\documentclass{iaus}
\usepackage{graphicx}

\title[Star Formation in the LMC] 
{Star Formation in the LMC: Gravitational Instability and Dynamical Triggering}

\author[Chu, Gruendl, \& Yang]   
{You-Hua Chu$^1$,
 Robert A. Gruendl$^1$ \break \and Chao-Chin Yang$^1$}

\affiliation{$^1$Astronomy Department, University of Illinois,
1002 W. Green Street, Urbana, IL 61801, USA \break email: chu@astro.uiuc.edu,
gruendl@astro.uiuc.edu, cyang8@astro.uiuc.edu}

\pubyear{2006}
\volume{237}  
\pagerange{}
\date{?? and in revised form ??}
\jname{Triggered Star Formation in a Turbulent Medium}
\editors{B. G. Elmegreen \& J. Palous, eds.}
\begin{document}

\maketitle

\begin{abstract}

Evidence for triggered star formation is difficult to establish
because energy feedback from massive stars tend to erase the
interstellar conditions that led to the star formation.
Young stellar objects (YSOs) mark sites of {\it current} star 
formation whose ambient conditions have not been significantly
altered.  {\it Spitzer} observations of the Large Magellanic
Cloud (LMC) effectively reveal massive YSOs.  The inventory
of massive YSOs, in conjunction with surveys of interstellar
medium, allows us to examine the conditions for star formation:
spontaneous or triggered.  We examine the relationship between
star formation and gravitational instability on a global scale,
and we present evidence of triggered star formation on local
scales in the LMC.

\end{abstract}

\firstsection 
\section{Introduction}

The star formation process is intertwined with the evolution 
of the interstellar medium (ISM).
After the onset of a burst of star formation, massive stars
photoionize the ambient ISM into an H\,II region, and 
subsequently energize the ambient medium via fast stellar 
winds and supernova ejecta to form a superbubble.
The expansion of H\,II regions and superbubbles can trigger 
further star formation.  
If triggered star formation continues at a high level over an
extended period of time, $>$10 Myr, a kpc-sized supergiant shell 
may be produced.
On the other hand, if a superbubble does not trigger a significant 
level of star formation, after the O stars have exploded and the 
remaining stars have dispersed, the superbubble will recombine into an
H\,I shell with no obvious concentration of stars within its boundary.

To illustrate {\it triggered} star formation, a causal relationship 
between the prenatal interstellar conditions and the formation of
stars need to be convincingly established.  
Main-sequence and evolved massive stars trace star formation in the
past few Myr, but their prenatal interstellar conditions have been 
significantly altered by the injection of stellar energies.
Massive young stellar objects (YSOs), having a short lifetime, trace 
the current star formation, while their prenatal interstellar 
conditions are still intact; therefore, massive YSOs and their
environment can be used to investigate triggered star formation.

\begin{figure}[t]
\begin{center}
\includegraphics[width=0.6\textwidth]{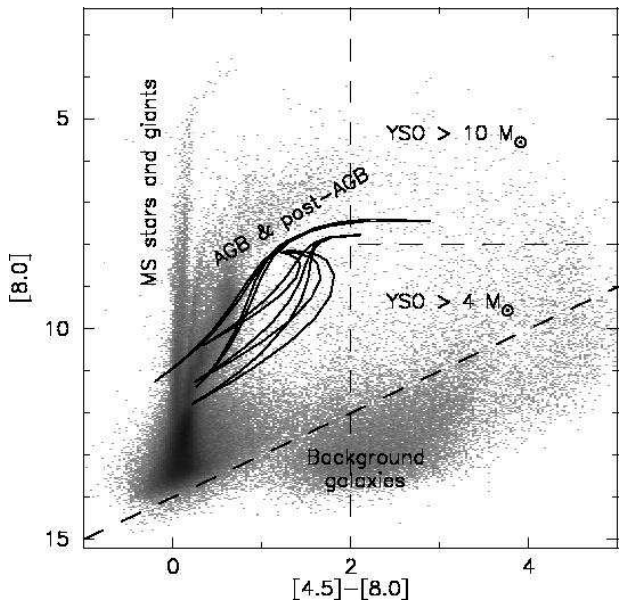}
\end{center}
\caption{Color-magnitude diagram of all point sources in the LMC.
Normal stars are at [4.5]$-$[8.0] $\sim$ 0.  The curves illustrate
locations of O-rich and C-rich AGB stars from Groenewegen (2006).  
The vertical dashed line roughly separate normal and AGB stars 
(to the left) and protostars (to the right).  The tilted dashed 
line marks a rough upper boundary of background galaxies determined
from the SWIRE data (Evans et al.\ 2005).}
\end{figure}

We have chosen the Large Magellanic Cloud (LMC) to investigate 
the star formation processes because of the following advantages:
(1) The LMC is at a small distance, 50 kpc, so that stars can be
resolved and the ISM can be mapped with high linear resolution, 
and it has a nearly face-on orientation so that confusion along 
the line-of-sight is minimal.
(2) The ISM of the LMC has been well mapped - MCELS survey of 
ionized gas (Smith 1999), ATCA+Parkes survey of the 
H\,I gas (Kim et al.\ 2003), and NANTEN survey of molecular
gas (Fukui et al.\ 2001). 
(3) A complete inventory of massive YSOs can be obtained from
the {\it Spitzer} IRAC and MIPS survey of the LMC (Meixner
et al.\ 2006).

We have identified point sources in the {\it Spitzer} IRAC 
and 24$\mu$m MIPS images and made photometric measurements.
Color-color and color-magnitude diagrams are used in conjunction 
with spectral energy distributions (SEDs) to diagnose YSO
candidates from these point sources.
Figure 1 shows an example of color-magnitude diagram in the IRAC
4.5 and 8.0 $\mu$m bands.
The main-sequence stars and giants have zero colors, [4.5]-[8.0]
$\sim$ 0.
Contaminations of AGB and post-AGB stars and background galaxies
can be largely avoided by employing appropriate color or brightness
cutoffs (Groenewegen 2006; Evans et al.\ 2005).
Known AGB stars and planetary nebulae are removed from the source
list.
The final YSO candidates are located in the upper right wedge in 
Figure 1.

While {\it Spitzer} observations allow us to identify YSOs in 
the LMC, the angular resolution severely limits our ability to
determine the YSOs' physical properties in detail.  
Figure 2 illustrates this problem in the H\,II region N11B.
The bright source identified at 8 $\mu$m has a SED consistent 
with a YSO with a substantial envelope. However, a high-resolution
2.1 $\mu$m image taken with the ISPI on the CTIO Blanco 4m telescope
shows at least three sources within the point-spread-function
of the 8 $\mu$m source; furthermore, a {\it Hubble Space Telescope}
({\it HST}) ACS/WFC H$\alpha$ image shows multiple sources within
a dust pillar whose surface is ionized.  It is not clear how much
nebular emission contributes to the 8 $\mu$m source and whether 
one or multiple YSOs are present.

The YSOs identified from {\it Spitzer} observations clearly do
not provide unambiguous information about their masses and
multiplicity.  They are nevertheless excellent markers of sites of
current star formation and allow us to examine the star formation 
process in the LMC both on global and local scales.

\begin{figure}[t]
\begin{center}
\includegraphics[width=0.8\textwidth]{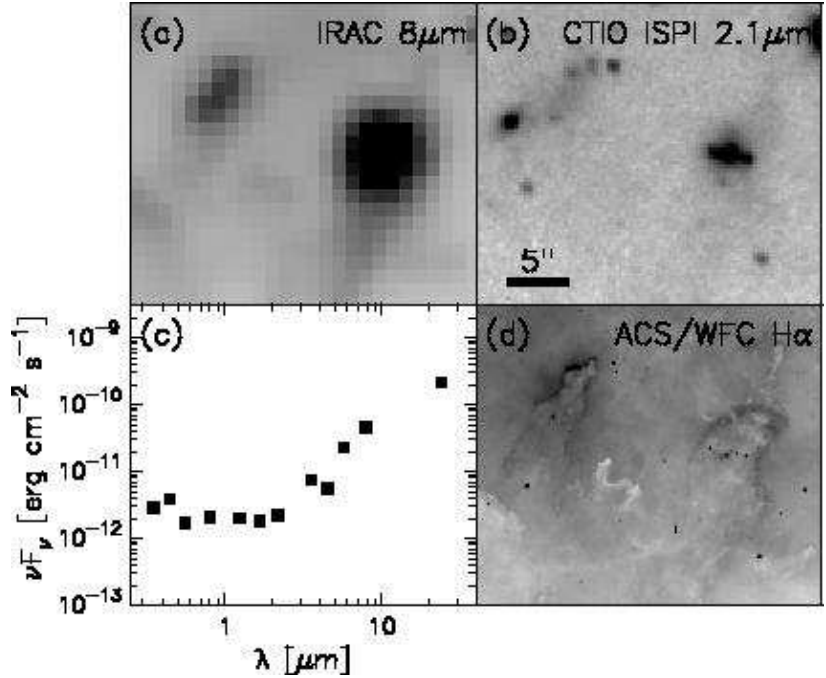}
\end{center}
\caption{(a) {\it Spitzer} IRAC 8 $\mu$m image of a region in N11B.
(b) CTIO Blanco 4m ISPI image of the same region at 2.1 $\mu$m.
(c) The SED of the brightest source in (a).
(d) {\it HST} ACS/WFC H$\alpha$ image of the same region.
}
\end{figure}

\begin{figure}[ht]
  \centering{
\includegraphics[width=0.49\textwidth]{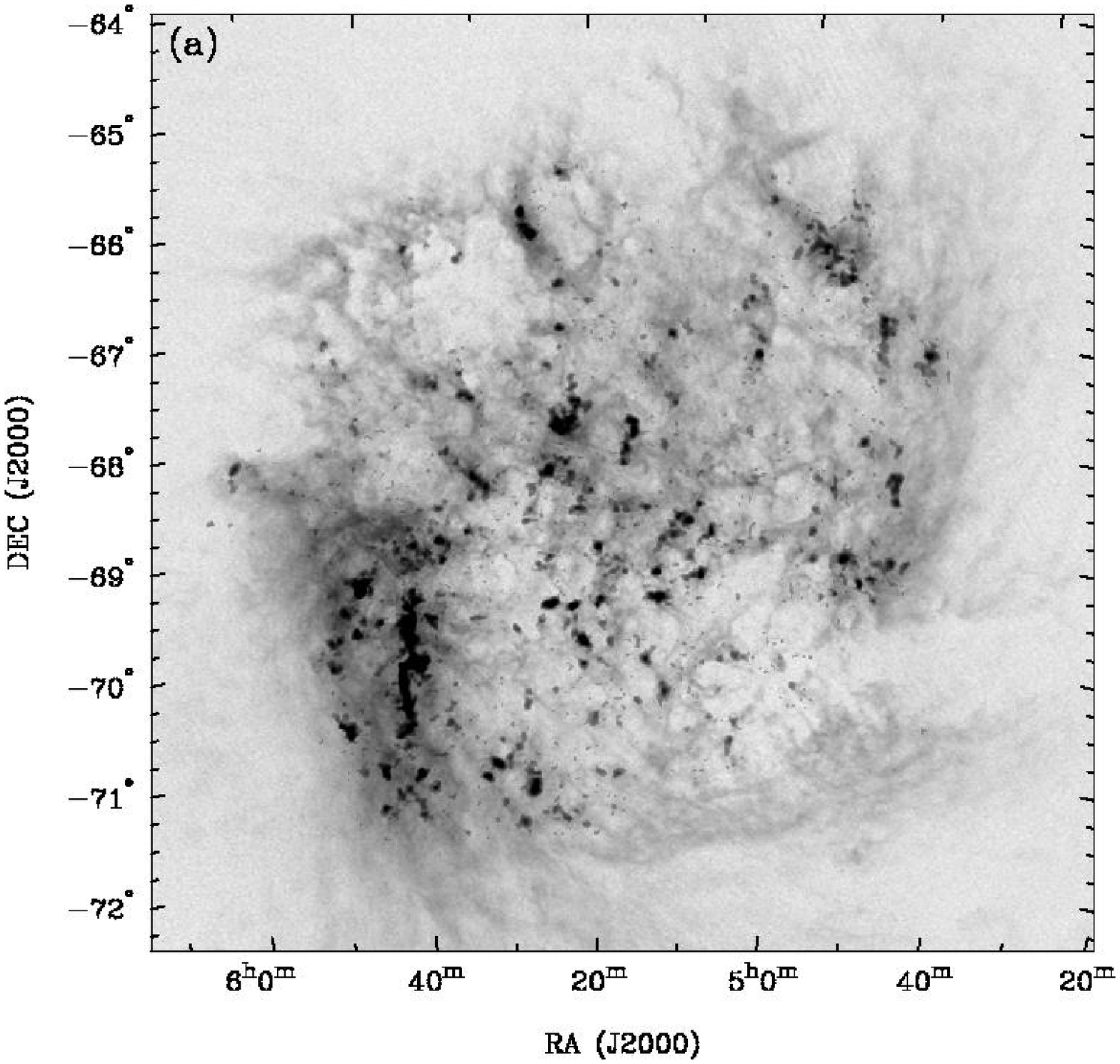}
\includegraphics[width=0.49\textwidth]{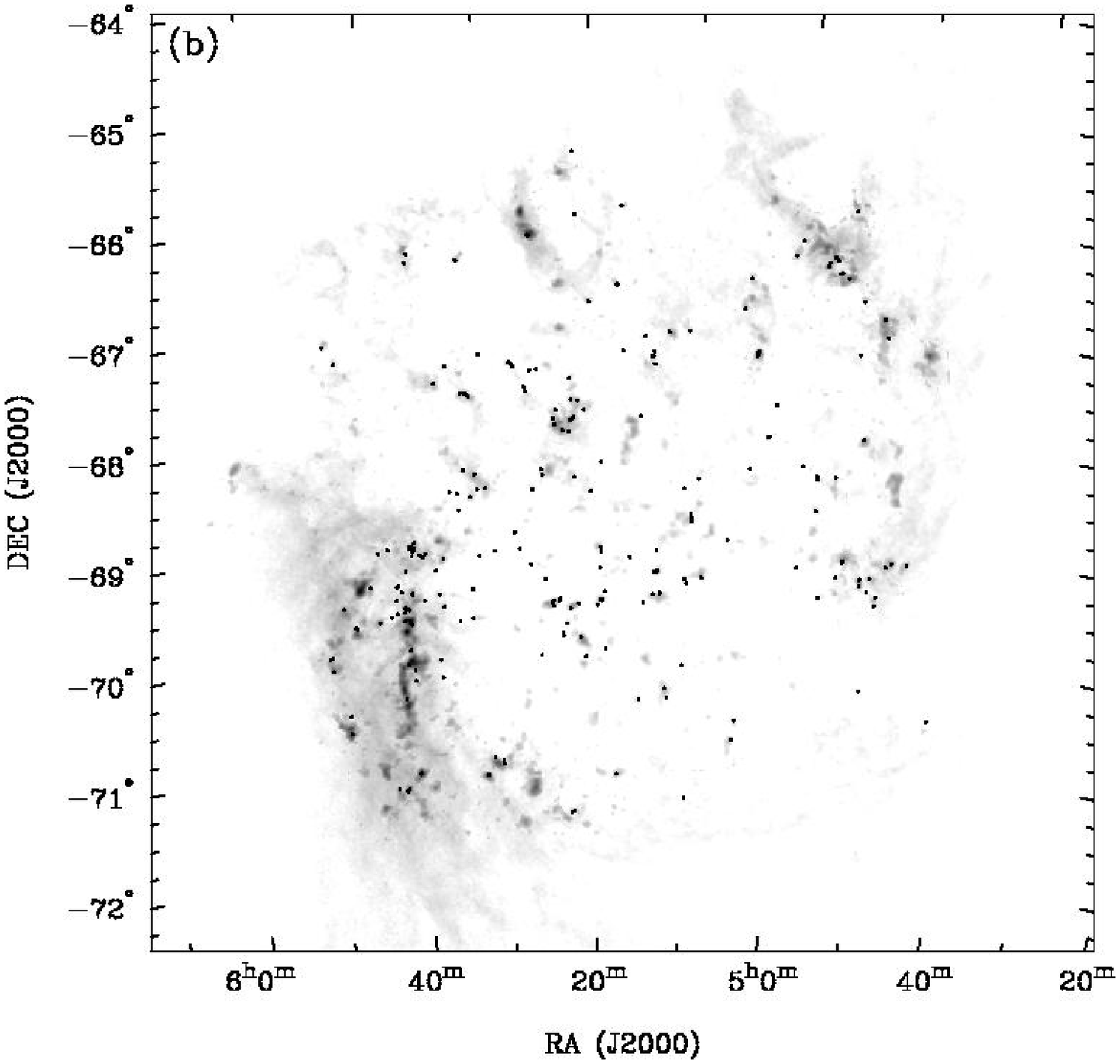}
\includegraphics[width=0.49\textwidth]{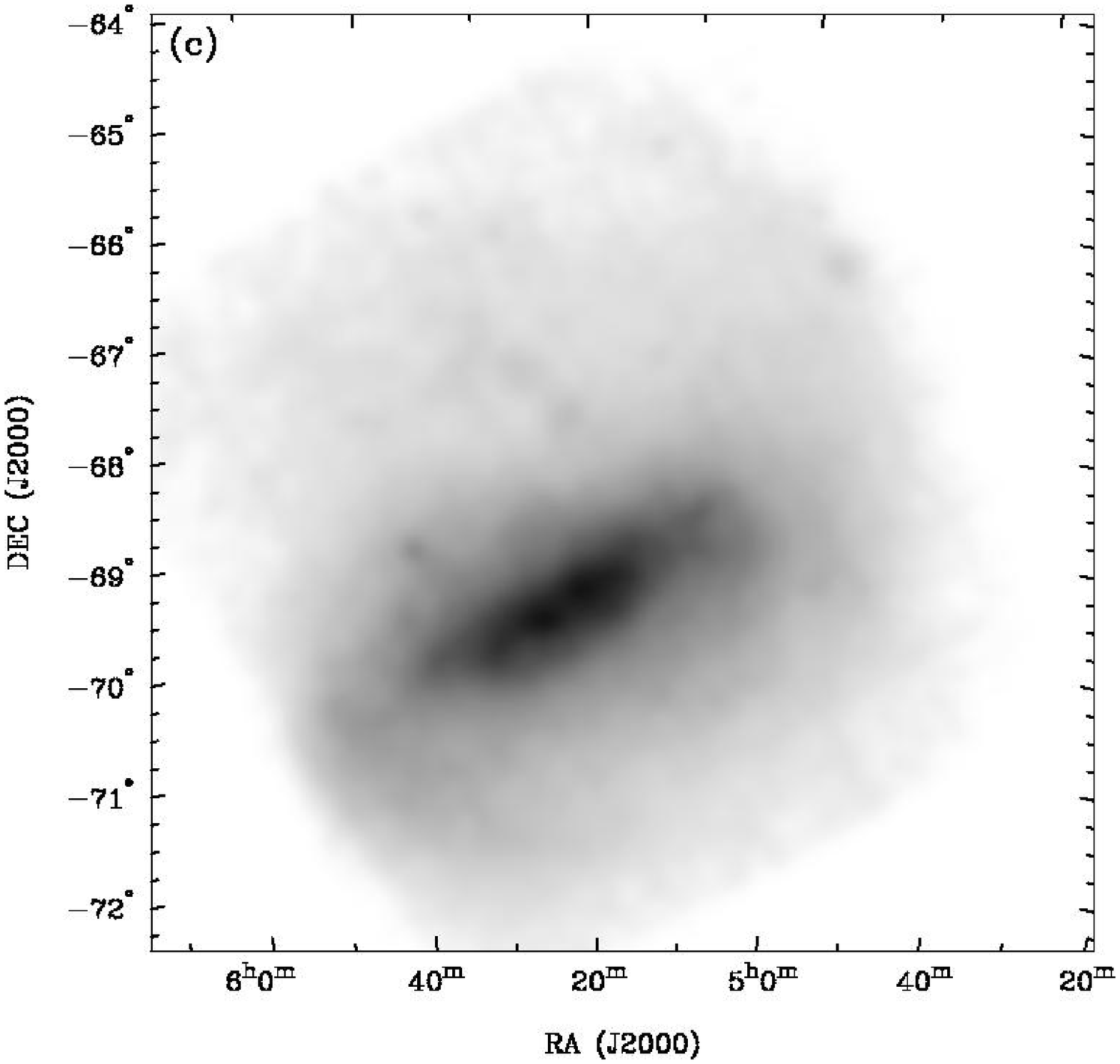}
\includegraphics[width=0.49\textwidth]{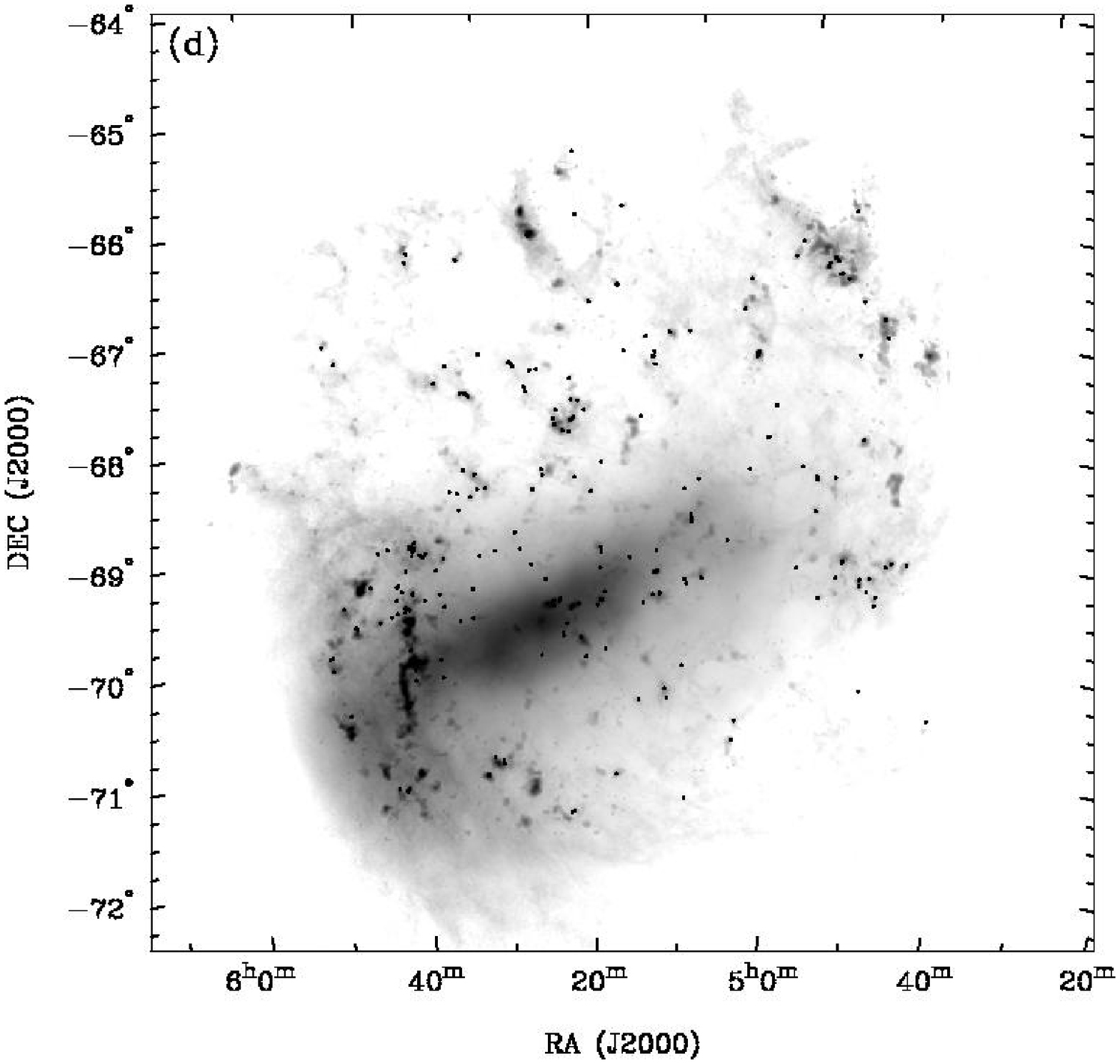}
  \caption{(a) Total gas surface density map of the LMC.
(b) The $Q_g$ map.  The unstable regions, with
    $Q_g < 1$, are shaded, and the darkness increases
    with the degree of instability.  YSO candidates
    are marked.
(c) Total stellar surface density map of the LMC, derived
    from the {\it Spitzer} 3.6 $\mu$m observations.
(d) The $Q_{sg}$ map. The unstable regions, with
    $Q_{sg} < 1$, are shaded, and the darkness increases
    with the degree of instability. YSO candidates are
    marked.    }}
  \label{fig:star}
\end{figure}

\section{Star Formation on a Global Scale}

Star formation activity is seen throughout the disk of the LMC.
We will examine the relationship between the distribution of
star formation and the gravitational instability on a global 
scale.
In the first case, we consider only a thin gaseous disk, and
in the second case we add the stellar contribution to the disk.

A thin, differentially rotating gaseous disk is stable to 
axisymmetric perturbations, if 
  $ Q_g \equiv \frac{\kappa c_g}{\pi G\Sigma_g} > 1$,
where $\Sigma_g$ and $c_g$ are the surface density and the sound
speed of the gas, and $\kappa$ is the epicycle frequency 
(Binney \& Tremaine 1987).
The total gas surface density map of the LMC (Fig.\ 3a) is derived 
from the the ATCA+Parkes H\,I survey by Kim et al.\ (2003) and the 
NANTEN CO survey by Fukui et al.\ (2001). 
(The total gas mass is $\sim6.5\times10^8~M_\odot$.)
The epicycle frequency as a function of radial distance is calculated
from the H\,I rotation curve (Kim et al. 1998).
The sound speed of gas is assumed to be 5 km~s$^{-1}$.
The resultant gravitational instability map of the LMC is
displayed in Figure 3b, where shaded regions are unstable, with
$ Q_g < 1$, and the darkness of the shade increases with the
degree of instability.
The Spitzer sample of massive YSO candidates are also marked
in Figure 3b.
It is evident that only about 1/2 of the YSO candidates are 
located in gravitationally unstable regions.

To add stellar contributions to the gravitational instability, we 
follow Rafikov's (2001) treatment of a disk galaxy consisting of a
collisional gas disk and a collisionless stellar disk.  The instability
condition becomes 
\begin{equation} \label{E:Qsg}
  \frac{1}{Q_{sg}}
  \equiv \frac{2}{Q_s}\frac{1}{q}\left[1 - e^{-q^2}I_0\left(q^2\right)\right] +
         \frac{2}{Q_g}\frac{1}{q}R\frac{q}{1 + q^2 R^2}
  > 1,
\end{equation}
where $Q_s \equiv \frac{\kappa\sigma_s}{\pi G\Sigma_s}$ with
$\Sigma_s$ and $\sigma_s$ being the surface density and the
radial velocity dispersion of the stars, $I_0$ is the Bessel function 
of order zero, $R \equiv \frac{c_g}{\sigma_s}$, and 
$q \equiv \frac{k\sigma_s}{\kappa}$ with $k$ being the wavenumber of the 
axisymmetric perturbations.

To estimate $\Sigma_s$, we use a normalized distribution of the 3.6 $\mu$m
sources detected in the {\it Spitzer} survey of the LMC (Gruendl et al.\
2007, in prep) and a total stellar mass of $2\times10^9~M_\odot$ 
(Kim et al.\ 1998).
Note that the 3.6 $\mu$m point sources, rather than the direct image,
are used to avoid the contamination of diffuse emission from dust, and
that each source is weighted by its brightness in the calculation.
The resulting stellar surface density map is shown in Figure 3c.
We adopt a stellar radial velocity dispersion of 15 km~s$^{-1}$,
and use the same $Q_g$ described above.
Finally, we calculate the value of $Q_{sg}$ at each pixel by finding
its global minimum as a function of the wavenumber $k$ (see Eq.\ \ref{E:Qsg}), 
following the same approach used by Jog (1996) and Rafikov (2001).
The resultant gravitational instability map of the LMC is presented
in Figure 3d, where shaded regions are unstable, with 
$ Q_{sg} < 1$, and the darkness of the shade increases with the
degree of instability.
The vast majority of the massive YSO candidates are located within
gravitationally unstable regions, in sharp contrast to the first case 
that includes only the gas disk.

From our analysis of the gravitational instability of the LMC, we
conclude that the stellar disk's contribution cannot be ignored.
More importantly, the current star formation appears to occur mostly
in regions that are unstable against perturbation, implying that
star formation can be triggered more easily in the gravitationally
unstable regions.

\section{Star Formation on Local Scales}

Triggered star formation is commonly seen in the LMC at
various scales.  Below we give three such examples.

\underline{Star formation triggered by H\,II region and enriched
by supernova ejecta} -- N63 shown in Figure 4.
The H\,II region N63 is ionized by the OB association LH\,83.  
A young supernova remnant (SNR), N63A,
has been identified in the H\,II region, as delineated by 
the X-ray emission.  {\it Spitzer} IRAC 8 $\mu$m 
image shows prominent current star formation on the northeastern 
rim of the H\,II region, not in contact with the SNR.

\underline{Star formation triggered by photo-implosion of dust globules
in a superbubble} -- N51D shown in Figure 5 (Chu et al.\ 2005).
Three YSOs projected within the superbubble N51D are found to be
coincident with dust globules.
The thermal pressure of the warm photoionized gas on the surface of 
the dust globules is several times higher than those of the hot gas in 
the superbubble interior and the cold molecular gas in the dust globules,
indicating that the dust globules are compressed by their high
external pressure raised by photoionization and the associated heating.

\underline{Star formation triggered by collision of interstellar shells}
-- supergiant shells LMC-4 and LMC-5 shown in Figure 6.
These two supergiant shells are expanding into each other, and the
colliding region shows a bright ridge of H\,I and molecular clouds.
On-going, active star formation is seen in the colliding region.

\section{Conclusions}\label{sec:concl}

The LMC is an excellent site to study star formation processes.
Our investigation suggests that global-scale gravitational instabilities 
prepare the conditions for star formation, but dynamical triggering 
on local scales determines where to set off the star formation and 
complete the job.

\begin{acknowledgments}
This research is supported by the {\it Spitzer} grant JPL 1264494.
\end{acknowledgments}

\newpage

\begin{figure}
\begin{center}
\includegraphics[width=0.8\textwidth]{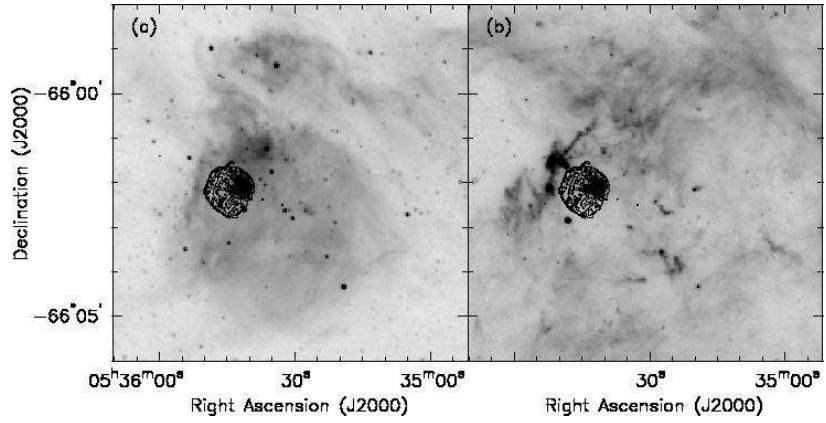}
\end{center}
\caption{(a) H$\alpha$ image of N63.  Overplotted are X-ray 
contours extracted from {\it Chandra} observations to show 
the spatial extent of the young supernova remnant N63A.
(b) {\it Spitzer} IRAC 8 $\mu$m image of N63.  YSO candidates
are detected along the northeastern rim of the H\,II region.
The star formation is triggered by the expansion of the H\,II
region, but the YSOs and their planetary disks will be enriched 
by the supernova ejecta.
}
\end{figure}

\begin{figure}
\begin{center}
\includegraphics[width=0.95\textwidth]{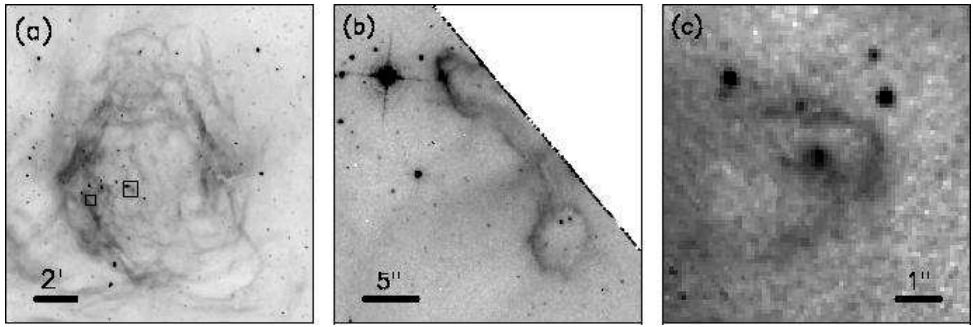}
\end{center}
\caption{(a) H$\alpha$ image of the superbubble N51D.  (b) \& (c) 
{\it HST} WFPC2 H$\alpha$ images of dust globules that host YSOs.  
The locations of these fields are marked by the two boxes in (a).
}
\end{figure}

\begin{figure}
\begin{center}
\includegraphics[width=0.45\textwidth]{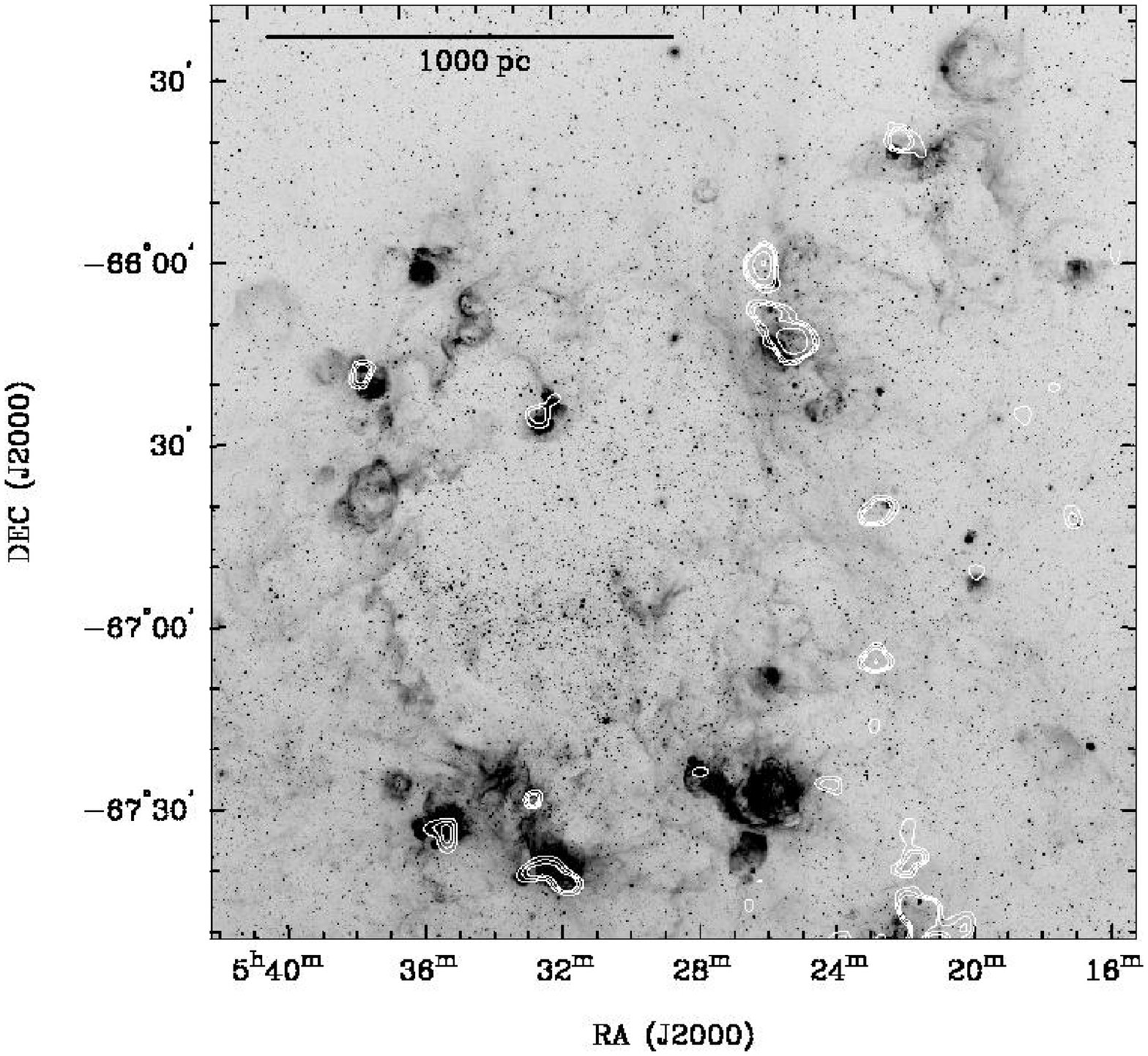}
\includegraphics[width=0.45\textwidth]{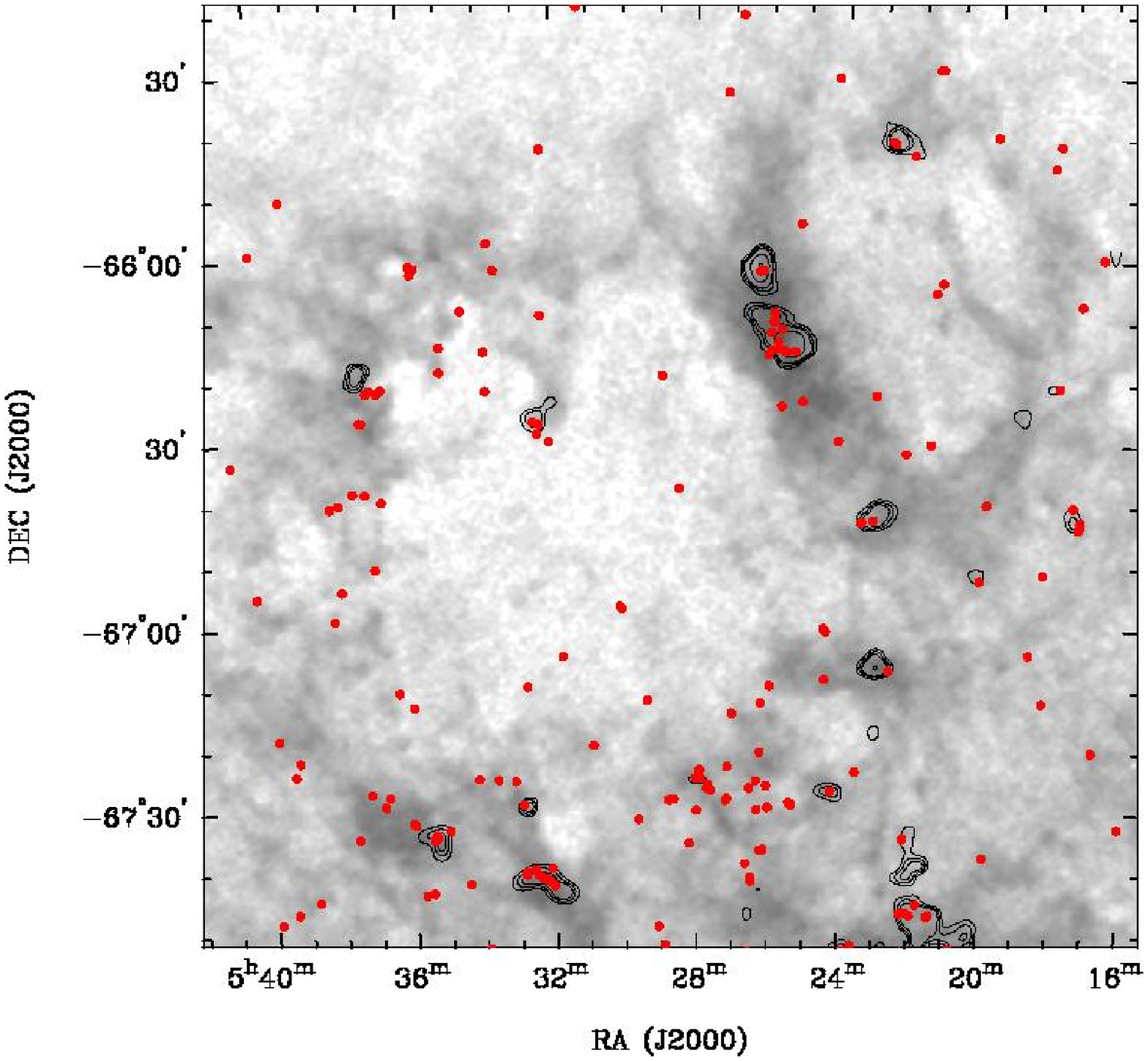}
\end{center}
\caption{{\it Left:} H$\alpha$ image of the supergiant shells LMC-4 and LMC-5
superposed with CO contours (Fukui et al.\ 2001). 
{\it Right:} ATCA+Parkes H\,I map of LMC-4 and LMC-5 in greay-scale.  
Overplotted are CO contours and positions of YSOs. Star formation is 
enhanced in the collision zone between the two supergiant shells.
}
\end{figure}

\end{document}